\documentclass[12pt]{article}
\usepackage{cite}
\usepackage{CJK}
\usepackage{amssymb}
\usepackage{amsmath}
\usepackage{enumerate}
\usepackage{graphicx}
\usepackage{epstopdf}
\usepackage{amsfonts}
\usepackage{url}
\usepackage{bm}
\usepackage{mathdots}
\usepackage{rotating}
\usepackage{pgfplots}
\usepackage{floatrow}
\usepackage{pifont}
\usepackage{epsfig,subfigure,dsfont,amsthm,amsbsy,mathrsfs,amscd}

\def\be{\begin{equation}}
\def\ee{\end{equation}}
\def\ba{\begin{array}}
\def\ea{\end{array}}
\input amssym.def

\newcommand\btd{\raise 2pt \hbox{$\hat\bigtriangledown$}\hskip 1.5pt}
\newcommand\bt{\raise 2pt \hbox{$\bigtriangledown$}\hskip 1.5pt}
\def\qed{\hfill \vrule height7pt width 7pt depth 0pt}

\begin{document}
 \title{\Large {\bf Distinguishing multipartite orthogonal product states by LOCC with entanglement as a resource }}
\author{Hai-Quan Li$^{1}$, Naihuan Jing$^{1,2}$$^\ast$  \& Xi-Lin Tang$^{1}$\\[10pt]
 {\footnotesize  {\small $^1$Department of Mathematics,
 South China University of Technology, Guangzhou
510640, P.R.China}} \\
 {\footnotesize  {\small $^2$Department of Mathematics, North Carolina State University, Raleigh, NC 27695, USA}} \\
}
\date{}

\maketitle
\noindent{\bf Abstract}
 Recently, the method that using an entanglement as a resource to distinguish orthogonal product states by local operations and classical communication (LOCC) has brought into focus. Zhang et al.
 presented protocols which use an entanglement to distinguish  some classes of orthogonal product states in $\mathbb{C}^m\otimes \mathbb{C}^n$\cite{Zhang016}.
 In this paper, we mainly study the local
  distinguishability of multipartite product states. For the class of locally indistinguishable multipartite product states constructed by Wang et al. in \cite{Wang17}, we present a protocol that distinguishes perfectly these quantum states by LOCC using an entangled state as a resource for implementing quantum measurements.


\section{Introduction}

\bigskip

In quantum information theory, the relationship between quantum nonlocality and quantum entanglement has received considerable attention in the last several decades due to their deep connections \cite{Ghosh01,Ghosh04,Hayashi06}. However, Bennett et al. found nine states without entanglement in  $\mathbb{C}^3\otimes\mathbb{C}^3$, which cannot be distinguished perfectly by local operations and classical communication (LOCC) \cite{Ben99}. This latter interesting phenomenon is called ``nonlocality without entanglement'', i.e., the local indistinguishability of mutually orthogonal product states by LOCC, and it has since attracted much attention \cite{Walgate02,Horodecki03,De04,Nathanson05,Niset06,Feng09,Duan10,Childs13,Zhang14,Wang15,Zhang15,Zhang16,Xu15,Xu16,Wang17,ZhangX17,Zhang17}.
These developments provided a better understanding of nonlocality without entanglement.

In $2008$, using an entanglement, Cohen perfectly distinguished certain classes of unextendible product bases (UPB) by LOCC in $\mathbb{C}^m\otimes\mathbb{C}^n$ \cite{Cohen07}. His method using entanglement as a resource to distinguish the quantum states showed
that entanglement is a valuable resource for quantum information. This is parallel to well-known theoretical applications of entanglement
such as in quantum information processing, quantum cryptography\cite{Ekert91, Gisin02}, quantum teleportation\cite{Karlsson98, Kim01}, and quantum secure direct communication \cite{Wang05, Tian08, Yang11}.  It is thus interesting to ask what entanglement resources are necessary and sufficient for distinguishing indistinguishable quantum states with LOCC.

In $2016$, Zhang et al. have shown that a $\mathbb{C}^2\otimes\mathbb{C}^2$ maximally entangled
states is sufficient to perfectly distinguish certain classes by LOCC in $\mathbb{C}^m\otimes\mathbb{C}^n$ \cite{Zhang016}
and have raised the question for a multipartite system. We believe that any class of locally indistinguishable orthogonal product states can be perfectly distinguished by LOCC with enough entanglements as resources.
On the other hand, Bandyopadhyay et al. have proved that there is no entangled state as a universal resource for local state discrimination in multipartite systems \cite{Bandyopadhyay16}. This result says that distinguishability of multipartite orthogonal product states
has to be dealt individually according to the system. 

The phenomenon of ``nonlocality without entanglement'' in multipartite quantum systems has also been studied in \cite{Niset06,Xu16,Wang17,Zhang17}.
Niset et al. constructed a set of locally indistinguishable multipartite orthogonal product bases in $\mathbb{C}^{d_1}\otimes\mathbb{C}^{d_2}\otimes\cdots\otimes\mathbb{C}^{d_{n-1}}\otimes\mathbb{C}^{d_{n}} (d_{i}\geq n-1)$\cite{Niset06}.
Xu et al. constructed two different classes of locally indistinguishable multipartite orthogonal product states in any multipartite quantum system \cite{Xu16}. In addition, Wang et al. constructed a set of LOCC indistinguishable multipartite orthogonal product states by using a set of locally indistinguishable bipartite orthogonal product states \cite{Wang17}. Recently, Zhang et al.\cite{Zhang17} gave a general construction of locally indistinguishable multipartite orthogonal product states. It is interesting to see
how to use entanglement as a resource to distinguish  multipartite orthogonal product states by LOCC.

In this paper, we consider the set of LOCC indistinguishable multipartite product states constructed in \cite{Wang17}
and show that they can be indeed distinguished by LOCC with entanglement as a resource. 
It is shown by separating the problem into the cases of even and odd partite states, 
and each case is solved by considering the lower rank cases.
For the even case, the local distinguishability of a class of orthogonal product states is considered in the bipartite system.
For the odd case, this is dealt with by first establishing the result in tripartite systems.

\bigskip
\section{\bf Local distinguishability of multipartite product states}
Let $\{|i\rangle\}_{i=1}^{d}$ be the standard orthonormal basis in $\mathbb{C}^d$.
A pure state $|\psi\rangle$ of $\mathbb{C}^{d}\otimes\mathbb{C}^{d'}(d' > d)$
is said to be maximally entangled if for any orthogonal 
basis ${|i_{A}\rangle}$ of the subsystem $A$, there exists an orthogonal basis ${|i_{B}\rangle}$ of the subsystem $B$ such that $|\psi\rangle$ can be written as $|\psi\rangle=\frac{1}{\sqrt{d}}\sum\limits_{i=1}^{d}|i_{A}i_{B}\rangle$\cite{Li12}.

In the following, we shall study local distinguishability of multipartite product states in even-partite and odd-partite cases respectively.
\subsection{Local distinguishability for even-partite case}

In this subsection, we first consider local distinguishability of orthogonal product states in a bipartite system.
We claim that any class of locally indistinguishable orthogonal product states can be perfectly distinguished by LOCC with enough entanglements as resources.

In $\mathbb{C}^m\otimes\mathbb{C}^n(4\leq m\leq n)$, the following $2n-1$ orthogonal product states are LOCC indistinguishable\cite{Wang17}.
\begin{equation}
\begin{split}
&|\alpha\pm \beta\rangle=\frac{1}{\sqrt{2}}(|\alpha\rangle\pm|\beta\rangle),0\leq\alpha\leq\beta,\\
&|\phi_1\rangle=(|1\rangle_A+|2\rangle_A+\cdots+|m\rangle_A)(|1\rangle_B+|2\rangle_B+\cdots+|n\rangle_B), \\
&|\phi_i\rangle=|i\rangle_A|1-i\rangle_B,  \ \ i=2, 3, \ldots, m, \\
&|\phi_{m+1}\rangle=|1-m\rangle_A|2\rangle_B, \\
&|\phi_{m+j-1}\rangle=|1-(j-1)\rangle_A|j\rangle_B,  \ \ j=3, 4, \ldots, m, \\
&|\phi_{m+l-1}\rangle=|1-2\rangle_A|l\rangle_B,  \ \ l=m+1,  m+2, \ldots,  n, \\
&|\phi_{m+n}\rangle=|m\rangle_A|3-(m+1)\rangle_B, \\
&|\phi_{n+s}\rangle=|m-1\rangle_A|s-(s+1)\rangle_B, \\
&|\phi_{n+t}\rangle=|m\rangle_A|t-(t+1)\rangle_B, \\
&s=m+2k-1, \,  t=m+2k, \,  k=1, 2, \ldots, \lfloor\frac{n-m}{2}\rfloor.
\end{split}
\end{equation}

Using the method presented by Cohen\cite{Cohen07}, we first need to add two ancillary systems $a$ and $b$ by sharing an entangle state $|\psi\rangle_{ab}$. Now the system $a$ and the system $A$ are both held by Alice, the system $b$ and the system $B$ are both held by Bob, i.e., Alice and Bob control $aA$ system and
$bB$ system respectively. Then, Bob measures $\{\emph{B}_i\}$ and performs one of the projectors on $bB$. The result is denoted by an operation $\emph{B}_1(|\omega_i\rangle_{AB}\otimes|\psi\rangle_{ab})$, where $|\omega_i\rangle_{AB}$ is the considered state. Finally, Alice and Bob can proceed from here to distinguish the states using only LOCC.

\bigskip
\noindent{\bf Theorem 1} In $\mathbb{C}^m\otimes\mathbb{C}^n(4\leq m\leq n)$, a $\mathbb{C}^n\otimes\mathbb{C}^n$ maximally entangled state is sufficient to perfectly distinguish the $2n-1$ orthogonal product states $(1)$ by LOCC.

\bigskip
\noindent{ \emph{Proof:}}
Let $|\psi\rangle_{ab}$ be a $\mathbb{C}^n\otimes\mathbb{C}^n$ maximally entangled state, $$|\psi\rangle_{ab}=\frac{1}{\sqrt{n}}\sum_{i=1}^n|ii\rangle_{ab}.$$
Alice and Bob share $|\psi\rangle_{ab}$, and then Bob performs the projector $\emph{B}_1=\sum_{i=1}^n|ii\rangle_{bB}\langle ii|$ on $bB$.

From $|\varphi_i\rangle=\emph{B}_1(|\phi_i\rangle_{AB}\otimes|\psi\rangle_{ab})$, we have

\begin{equation}
\begin{split}
&|\varphi_1\rangle=(|1\rangle_A+|2\rangle_A+\cdots+|m\rangle_A)(|1\rangle_B|11\rangle_{ab}+|2\rangle_B|22\rangle_{ab}+\cdots+|n\rangle_B|nn\rangle_{ab}), \\
&|\varphi_i\rangle=|i\rangle_A(|1\rangle_B|11\rangle_{ab}-|i\rangle_B|ii\rangle_{ab}), \ \ i=2, 3, \ldots, m, \\
&|\varphi_{m+1}\rangle=|1-m\rangle_A|2\rangle_B|22\rangle_{ab}, \\
&|\varphi_{m+j-1}\rangle=|1-(j-1)\rangle_A|j\rangle_B|jj\rangle_{ab},  \ \ j=3, 4, \ldots, m, \\
&|\varphi_{m+l-1}\rangle=|1-2\rangle_A|l\rangle_B|ll\rangle_{ab}, \ \ l=m+1,  m+2, \ldots,  n, \\
&|\varphi_{m+n}\rangle=|m\rangle_A(|3\rangle_B|33\rangle_{ab}-|m+1\rangle_B|(m+1)(m+1)\rangle_{ab}), \\
&|\varphi_{n+s}\rangle=|m-1\rangle_A(|s\rangle_B|ss\rangle_{ab}-|s+1\rangle_B|(s+1)(s+1)\rangle_{ab}), \\
&|\varphi_{n+t}\rangle=|m\rangle_A(|t\rangle_B|tt\rangle_{ab}-|t+1\rangle_B|(t+1)(t+1)\rangle_{ab}), \\
&s=m+2k-1, \,  t=m+2k, \,  k=1, 2, \ldots, \lfloor\frac{n-m}{2}\rfloor.
\end{split}
\end{equation}

To distinguish these states using LOCC, Alice makes a projective measurement with $2n-2$ outcomes. Considering the outcome $\emph{A}_{i-1}=|1\rangle_a\langle1|\otimes|i\rangle_A\langle i|+|i\rangle_a\langle i|\otimes|i\rangle_A\langle i|$, it leaves $|\varphi_{i}\rangle$ invariant and transforms $|\varphi_{1_{i}}\rangle$ to $|i\rangle_A(|1\rangle_{B}|11\rangle_{ab}+|i\rangle_{B}|ii\rangle_{ab})$ for $i=2, 3, \ldots, m$.
Since $|\varphi_{i}\rangle$ and $|\varphi_{1_{i}}\rangle$ are two mutually orthogonal states, Bob can easily distinguish the two states using the projector $\emph{B}_{(i-1)1, (i-1)2}=|1\rangle_b\langle1|\otimes|1\rangle_B\langle1|\pm|i\rangle_b\langle i|\otimes|i\rangle_B\langle i|$ for $i=2, 3, \ldots, m$.

Take the outcome $\emph{A}_m=|2\rangle_a\langle2|\otimes|1-m\rangle_A\langle1-m|$, the only remaining possibility is $|\varphi_{m+1}\rangle$, which has thus been successfully identified. In the same way, Alice can identify $|\varphi_{m+j-1}\rangle$
and $|\varphi_{m+l-1}\rangle$ by projector $\emph{A}_{m+j-2}=|j\rangle_a\langle j|\otimes|1-(j-1)\rangle_A\langle1-(j-1)|$ and $\emph{A}_{m+l-2}=|l\rangle_a\langle l|\otimes|1-2\rangle_A\langle1-2|$ ($j=3, 4, \ldots, m$, $l=m+1, m+2,\ldots, n$) respectively.

For the outcome $\emph{A}_{m+n-1}=|3\rangle_a\langle3|\otimes|m\rangle_A\langle m|+|(m+1)\rangle_a\langle(m+1)|\otimes|m\rangle_A\langle m|$, it leaves $|\varphi_{m+n}\rangle$ invariant and transforms $|\varphi_{1_{m+n}}\rangle$ to $|m\rangle_A(|3\rangle_{B}|33\rangle_{ab}+|m+1\rangle_{B}|(m+1)(m+1)\rangle_{ab})$. Bob can easily identify the two states, since any two orthogonal states can be distinguished. In the same way, $|\varphi_{n+s}\rangle$ and $|\varphi_{m+t}\rangle$ can be distinguished for $s=m+2k-1$ and $t=m+2k,$ where $k=1, 2, \ldots, \lfloor\frac{n-m}{2}\rfloor$.

Therefore, we have succeeded in distinguishing the states (1) by LOCC using a $\mathbb{C}^n\otimes\mathbb{C}^n$ maximally entangled state as a resource.\qed

\bigskip
\noindent{\bf Example 1 } In $\mathbb{C}^4\otimes\mathbb{C}^5$, a $\mathbb{C}^5\otimes\mathbb{C}^5$ maximally entangled state is sufficient to perfectly distinguish the following $9$ LOCC indistinguishable states by LOCC.

\begin{equation}
\begin{split}
&|\alpha\pm \beta\rangle=\frac{1}{\sqrt{2}}(|\alpha\rangle\pm|\beta\rangle),0\leq\alpha\leq\beta,\\
&|\phi_1\rangle=(|1\rangle_A+|2\rangle_A+\cdots+|4\rangle_A)(|1\rangle_B+|2\rangle_B+\cdots+|5\rangle_B), \\
&|\phi_2\rangle=|2\rangle_A|1-2\rangle_B,\ \ \ \ \ \ \ \ |\phi_3\rangle=|3\rangle_A|1-3\rangle_B, \\
&|\phi_4\rangle=|4\rangle_A|1-4\rangle_B, \ \ \ \ \ \ \ \  |\phi_5\rangle=|1-4\rangle_A|2\rangle_B, \\
&|\phi_6\rangle=|1-2\rangle_A|3\rangle_B, \ \ \ \ \ \ \ \  |\phi_7\rangle=|1-3\rangle_A|4\rangle_B, \\
&|\phi_8\rangle=|1-2\rangle_A|5\rangle_B, \ \ \ \ \ \ \ \ |\phi_9\rangle=|4\rangle_A|3-5\rangle_B. \\
\end{split}
\end{equation}

These states can be described more specifically by a box-diagram (Fig. 1), where all states are shown except $|\phi_1\rangle$, as the latter
 would
have covered the whole picture.

\bigskip
In fact, it follows from the proof of Theorem 1 that one uses the states
$$|\psi\rangle_{ab}=\frac{1}{\sqrt{5}}(|11\rangle_{ab}+|22\rangle_{ab}+|33\rangle_{ab}+|44\rangle_{ab}+|55\rangle_{ab}),$$ $$\emph{B}_1=|11\rangle_{bB}\langle11|+|22\rangle_{bB}\langle22|+|33\rangle_{bB}\langle33|+|44\rangle_{bB}\langle44|+|55\rangle_{bB}\langle55|$$
to perform $\emph{B}_1(|\phi_i\rangle_{AB}\otimes|\psi\rangle_{ab})$ and the results are the following states
\begin{equation}
\begin{split}
&|\varphi_1\rangle=(|1\rangle_A+|2\rangle_A+|3\rangle_A+|4\rangle_A)(|1\rangle_B|11\rangle_{ab}+|2\rangle_B|22\rangle_{ab}+|3\rangle_B|33\rangle_{ab}+|4\rangle_B|44\rangle_{ab} \\
&~~~~~~~~~+|5\rangle_B|55\rangle_{ab}),\\
&|\varphi_2\rangle=|2\rangle_A(|1\rangle_B|11\rangle_{ab}-|2\rangle_B|22\rangle_{ab}), \\
&|\varphi_3\rangle=|3\rangle_A(|1\rangle_B|11\rangle_{ab}-|3\rangle_B|33\rangle_{ab}), \\
&|\varphi_4\rangle=|4\rangle_A(|1\rangle_B|11\rangle_{ab}-|4\rangle_B|44\rangle_{ab}), \\
&|\varphi_5\rangle=|1-4\rangle_A|2\rangle_B|22\rangle_{ab}, \\
&|\varphi_6\rangle=|1-2\rangle_A|3\rangle_B|33\rangle_{ab}, \\
&|\varphi_7\rangle=|1-3\rangle_A|4\rangle_B|44\rangle_{ab}, \\
&|\varphi_8\rangle=|1-2\rangle_A|5\rangle_B|55\rangle_{ab}, \\
&|\varphi_9\rangle=|4\rangle_A(|3\rangle_B|33\rangle_{ab}-|5\rangle_B|55\rangle_{ab}),
\end{split}
\end{equation}

These states are depicted on Fig. 2, where the stopper state $|\varphi_1\rangle$ is not shown due to the same reason as above.

 \begin{figure}[h]
 \floatsetup{floatrowsep=qquad}
 \begin{floatrow}
 \ffigbox[\FBwidth]{\caption{The Tiles product states on a $\mathbb{C}^4\otimes\mathbb{C}^5$ system. }\label{bipartite_Example1}}{\includegraphics[width=0.36\textwidth,height=0.4\textwidth]{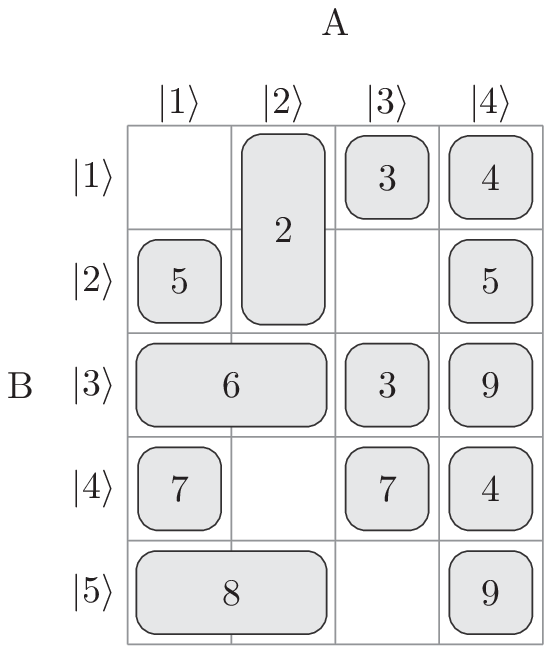}}

\ffigbox[\FBwidth]{\caption{The Tiles product states when Alice and Bob share a $\mathbb{C}^5\otimes\mathbb{C}^5$ MES and implementing quantum measurements.}\label{bipartite_Example2}}
{\includegraphics[width=0.36\textwidth,height=0.5\textwidth]{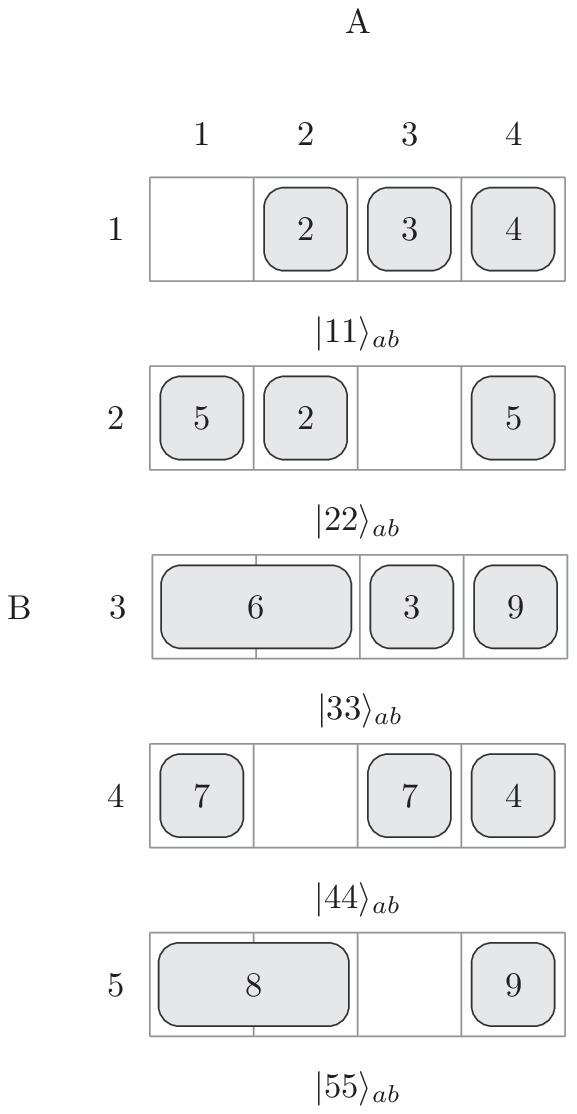}
}
 \end{floatrow}
\end{figure}

As an example, the state $|\phi_2\rangle$ is obtained as follows.
\begin{equation}
\begin{split}
|\varphi_2\rangle&=\emph{B}(|\phi_2\rangle_{AB}\otimes|\psi\rangle_{ab})\\
&=(|11\rangle_{bB}\langle11|+\cdots+|55\rangle_{bB}\langle55|)(|2\rangle_A|1-2\rangle_B
(|11\rangle_{ab}+|22\rangle_{ab}+\cdots+|55\rangle_{ab}))\\
&=(|11\rangle_{bB}\langle11|+\cdots+|55\rangle_{bB}\langle55|)(|2\rangle_A(|111\rangle_{abB}+|221\rangle_{abB}+\cdots+|551\rangle_{abB})\\
&~~~ -|2\rangle_A(|112\rangle_{abB}+|222\rangle_{abB}+\cdots+|552\rangle_{abB}))\\
&=|2\rangle_A(|1\rangle_B|11\rangle_{ab}-|2\rangle_B|22\rangle_{ab})
\end{split}
\end{equation}

Next we can distinguish the states easily as follows. Alice makes an eight-outcome projective measurement, and one
begins by considering the first outcome,
$\emph{A}_1=|1\rangle_a\langle1|\otimes|2\rangle_A\langle2|+|2\rangle_a\langle2|\otimes|2\rangle_A\langle2|$.
This leaves the state $|\varphi_{2}\rangle$ invariant and transforms $|\varphi_{1}\rangle$ to $|2\rangle_A(|1\rangle_B|11\rangle_{ab}+|2\rangle_B|22\rangle_{ab})$.
Then Bob uses the projectors $\emph{B}_{11, 12}=|1\rangle_b\langle1|\otimes|1\rangle_B\langle1|\pm|2\rangle_b\langle2|\otimes|2\rangle_B\langle2|$, which
can be easily identified to be $|\varphi_{1}\rangle$ and $|\varphi_{2}\rangle$ respectively. In the same way,  for outcomes $\emph{A}_2=|1\rangle_a\langle1|\otimes|3\rangle_A\langle3|+|3\rangle_a\langle3|\otimes|3\rangle_A\langle3|$ and
$\emph{A}_3=|1\rangle_a\langle1|\otimes|4\rangle_A\langle4|+|4\rangle_a\langle4|\otimes|4\rangle_A\langle4|$, Bob can identify $|\varphi_{3, 4}\rangle$ by
projectors $\emph{B}_{21, 22}=|1\rangle_b\langle1|\otimes|1\rangle_B\langle1|\pm|3\rangle_b\langle3|\otimes|3\rangle_B\langle3|$ and
$\emph{B}_{31, 32}=|1\rangle_b\langle1|\otimes|1\rangle_B\langle1|\pm|4\rangle_b\langle4|\otimes|4\rangle_B\langle4|$, respectively.

Consider the outcome $\emph{A}_4=|2\rangle_a\langle2|\otimes|1-4\rangle_A\langle1-4|$, the only remaining possibility is
$|\varphi_5\rangle$, which has thus been successfully identified. In the same way, Alice can identify $|\varphi_{6, 7, 8}\rangle$ by three
projectors $\emph{A}_5=|3\rangle_a\langle3|\otimes|1-2\rangle_A\langle1-2|$, $\emph{A}_6=|4\rangle_a\langle4|\otimes|1-3\rangle_A\langle1-3|$,
$\emph{A}_7=|5\rangle_a\langle5|\otimes|1-2\rangle_A\langle1-2|$, respectively.

For the last outcome $\emph{A}_8=|3\rangle_a\langle3|\otimes|4\rangle_A\langle4|+|5\rangle_a\langle5|\otimes|4\rangle_A\langle4|$, it leaves $|\varphi_{9}\rangle$ invariant and transforms $|\varphi_{1}\rangle$ to $|4\rangle_A(|3\rangle_B|33\rangle_{ab}+|5\rangle_B|55\rangle_{ab})$. Since $|\varphi_{1}\rangle$ and $|\varphi_{9}\rangle$ are two mutually orthogonal states, Bob can easily distinguish them 
by using the projectors $\emph{B}_{81, 82}=|3\rangle_b\langle3|\otimes|3\rangle_B\langle3|\pm|5\rangle_b\langle5|\otimes|5\rangle_B\langle5|$.

Thus the $9$ quantum states (3) have been perfectly distinguished using solely LOCC with a $\mathbb{C}^5\otimes\mathbb{C}^5$ maximally entangled state.\qed

\bigskip
Next we consider local distinguishability of orthogonal product states for the general even-partite case. Surprisingly, a set of locally indistinguishable multipartite product states can be constructed by using a set of locally indistinguishable bipartite orthogonal product states. Recall that
Wang et al.\cite{Wang17} have constructed $2(n_2+n_4+\cdots+n_{2k}-k)+1$ LOCC indistinguishable orthogonal product states in $\mathbb{C}^{n_1}\otimes\mathbb{C}^{n_2}\otimes\cdots\otimes\mathbb{C}^{n_{2k-1}}\otimes\mathbb{C}^{n_{2k}}~(4\leq n_1\leq n_2\leq\cdots\leq n_{2k-1}\leq n_{2k},  k\geq 2)$.

The set is given explicitly by
{\small
$$
\begin{array}{rcl}
  \mathcal{S}_1 & = & \{|\phi_{i_1}\rangle_1|11\rangle_2\cdots|11\rangle_k\  \big| \  i_1=1,2,\ldots,2n_2-2\}, \\
  \mathcal{S}_2 & = & \{|111\rangle_1|\phi_{i_2}\rangle_2|11\rangle_3\cdots|11\rangle_k  \big| \  i_2=1,2,\ldots,2n_4-2\}, \\
                & \vdots &                                                       \\
  \mathcal{S}_k & = & \{|111\rangle_1\cdots|11\rangle_{k-1}|\phi_{i_k}\rangle_k  \big| \  i_k=1,2,\ldots,2n_{2k}-2\}.
\end{array}
$$
}

Here the stopper state $|\phi\rangle=|\phi_1\rangle|\phi_2\rangle\ldots|\phi_{k}\rangle$, where $|\phi_s\rangle$ is the stopper state of the bipartite system
$\mathbb{C}^{n_{2s-1}}\otimes\mathbb{C}^{n_{2s}}$~($s = 1,2, \ldots, k$).

\bigskip
\noindent{\bf Theorem \ 2.}
In $\mathbb{C}^{n_1}\otimes\mathbb{C}^{n_2}\otimes\cdots\otimes\mathbb{C}^{n_{2k-1}}\otimes\mathbb{C}^{n_{2k}}$, it is sufficient to perfectly distinguish the set $\mathcal{S}$ of orthogonal product states by LOCC using entanglement as a resource.

\bigskip
\noindent{ \emph{Proof:}}
The set $\mathcal{S}$ of the $2(n_2+n_4+\cdots+n_{2k}-k)+1$ multipartite product states consists of the sets $\mathcal{S}_{i} ~(i=1,2,\ldots,k)$ and a stopper state.
If the states in each set $\mathcal{S}_i$ can be distinguished, then all the states in $\mathcal{S}$ can be distinguished successfully.

Let $\mathcal{H}$ = $\mathbb{C}^{n_1}\otimes\mathbb{C}^{n_2}\otimes\cdots\otimes\mathbb{C}^{n_{2k-1}}\otimes\mathbb{C}^{n_{2k}}(4\leq n_1\leq n_2\leq\cdots\leq n_{2k-1}\leq n_{2k})$. Suppose that each spatially separated observer $\mathcal{O}_i$~($i=1,2,\ldots,2k$) controls a subsystem of $\mathcal{H}$. For example, Alice controls the subsystem $\mathcal{H^{A}}$ and Bob controls the subsystem $\mathcal{H^{B}}$ in the bipartite system $\mathcal{H^{A}}\otimes\mathcal{H^{B}}$. We call the observers $\mathcal{O}_{2s-1}$ and $\mathcal{O}_{2s}$  the $s$-th Alice and the $s$-th Bob respectively. Here we can see that $\mathcal{H}$ consists of $k$ local subsystems $\mathbb{C}^{n_{2s-1}}\otimes\mathbb{C}^{n_{2s}}, s = 1,2, \ldots, k$. Let $|\psi\rangle$ = $\otimes_{s=1}^k|\psi\rangle_{s}$ be an entangle state in $\mathcal{H}$, where $|\psi\rangle_{s}$ is a $\mathbb{C}^{n_{2s}}\otimes\mathbb{C}^{n_{2s}}$ maximally entangled state for $s = 1,2, \ldots, k.$ The parties share the entangle state $|\psi\rangle$. That is, the $s$-th Alice and the $s$-th Bob share the maximally entangled state $|\psi\rangle_{s}$.

We denote the states $|\psi_{i_s}\rangle=|\phi_{i_s}\rangle_1|11\rangle_2\cdots|11\rangle_k$ ($i_s=1, 2, \ldots, 2n_{2s}-2$) and $|\psi_{2n_{2s}-1}\rangle=|\phi\rangle$. Because of the particularity of multipartite product states, the local distinguishability of the bipartite states $|\phi_{i_s}\rangle$~($i_s=1, 2, \ldots, 2n_{2s}-2$) and $|\phi_{s}\rangle$ is corresponding to the local distinguishability of the multipartite states $|\psi_{i_s}\rangle$~($i_s=1, 2, \ldots, 2n_{2s}-1$). By Theorem 1, since the $s$-th Alice and the $s$-th Bob share the maximally entangled state $|\psi\rangle_{s}$, the bipartite states $|\phi_{i_s}\rangle$, $i_s=1, 2, \ldots, 2n_{2s}-2$ and $|\phi_{s}\rangle$ can be distinguished with LOCC. Thus, the multipartite states $|\psi_{i_s}\rangle$, $i_s=1, 2, \ldots, 2n_{2s}-1$, also can be distinguished with LOCC.

Therefore, the $2(n_2+n_4+\cdots+n_{2k}-k)+1$ indistinguishable quantum states can be LOCC distinguished using the maximally entangled state $|\psi\rangle$ as a resource.\qed

\bigskip
\subsection{Local distinguishability for odd-partite case}

In this subsection, we study local distinguishability of orthogonal product states for odd-partite systems. We use
a concrete example to illustrate the idea, and then present the general method for the tripartite system. Based on tripartite systems, we then establish the general result for local distinguishability 
in any odd multipartite system.

In $\mathbb{C}^4\otimes\mathbb{C}^5\otimes\mathbb{C}^6$, Wang. et al.\cite{Wang17} have given $17$ LOCC indistinguishable orthogonal product states as follows

\begin{equation}
\begin{split}
&|\phi_1\rangle=|1+2+3+4\rangle_C|1+2+3+4+5\rangle_B|1+2+3+4+5+6\rangle_A,\\
&|\phi_2\rangle=|4\rangle_C|2\rangle_B|1-2\rangle_A,~~~~~~~~|\phi_{10}\rangle=|4\rangle_C|5\rangle_B|3-6\rangle_A,\\
&|\phi_3\rangle=|4\rangle_C|3\rangle_B|1-3\rangle_A,~~~~~~~~|\phi_{11}\rangle=|4\rangle_C|1-2\rangle_B|6\rangle_A,\\
&|\phi_4\rangle=|4\rangle_C|4\rangle_B|1-4\rangle_A,~~~~~~~~|\phi_{12}\rangle=|3\rangle_C|1-2\rangle_B|6\rangle_A,\\
&|\phi_5\rangle=|4\rangle_C|5\rangle_B|1-5\rangle_A,~~~~~~~~|\phi_{13}\rangle=|2\rangle_C|1-2\rangle_B|6\rangle_A,\\
&|\phi_6\rangle=|4\rangle_C|1-5\rangle_B|2\rangle_A,~~~~~~~~|\phi_{14}\rangle=|1\rangle_C|1-2\rangle_B|6\rangle_A,\\
&|\phi_7\rangle=|4\rangle_C|1-2\rangle_B|3\rangle_A,~~~~~~~~|\phi_{15}\rangle=|1-2\rangle_C|4\rangle_B|6\rangle_A,\\
&|\phi_8\rangle=|4\rangle_C|1-3\rangle_B|4\rangle_A,~~~~~~~~|\phi_{16}\rangle=|2-3\rangle_C|5\rangle_B|6\rangle_A,\\
&|\phi_9\rangle=|4\rangle_C|1-4\rangle_B|5\rangle_A,~~~~~~~~|\phi_{17}\rangle=|3-4\rangle_C|4\rangle_B|6\rangle_A.\\
\end{split}
\end{equation}

The LOCC indistinguishability of the $17$ states is derived from the fact that no party can go first, i.e., 
Alice and Bob cannot apply any nontrivial measurements, and the third party Charlie can only apply the trivial measurement.
By introducing a maximally entangled state shared by Alice and Bob, Alice use projectors such that
the three parties can perform measurements to distinguish the results by LOCC. The procedure is carried in two steps.

First Alice and Bob share the $\mathbb{C}^6\otimes\mathbb{C}^6$ maximally entangled state $$|\psi\rangle_{ab}=|11\rangle_{ab}+|22\rangle_{ab}+|33\rangle_{ab}+|44\rangle_{ab}+|55\rangle_{ab}+|66\rangle_{ab}.$$

Then Alice performs the projector $$\emph{A}_1=|11\rangle_{aA}\langle11|+|22\rangle_{aA}\langle22|+|33\rangle_{aA}\langle33|+|44\rangle_{aA}\langle44|
+|55\rangle_{aA}\langle55|+|66\rangle_{aA}\langle66|$$
to get the following states:
\begin{equation}
\begin{split}
&|\phi'_1\rangle=|1+2+3+4\rangle_C|1+2+3+4+5\rangle_B\sum_{i=1}^6|iii\rangle_{Aab},\\
&|\phi'_2\rangle=|4\rangle_C|2\rangle_B(|111\rangle_{Aab}-|222\rangle_{Aab}),\\
&|\phi'_3\rangle=|4\rangle_C|3\rangle_B(|111\rangle_{Aab}-|333\rangle_{Aab}),\\
&|\phi'_4\rangle=|4\rangle_C|4\rangle_B(|111\rangle_{Aab}-|444\rangle_{Aab}),\\
&|\phi'_5\rangle=|4\rangle_C|5\rangle_B(|111\rangle_{Aab}-|555\rangle_{Aab}),\\
&|\phi'_6\rangle=|4\rangle_C|1-5\rangle_B|222\rangle_{Aab},\\
&|\phi'_7\rangle=|4\rangle_C|1-2\rangle_B|333\rangle_{Aab},\\
&|\phi'_8\rangle=|4\rangle_C|1-3\rangle_B|444\rangle_{Aab},\\
&|\phi'_9\rangle=|4\rangle_C|1-4\rangle_B|555\rangle_{Aab},\\
&|\phi'_{10}\rangle=|4\rangle_C|5\rangle_B(|333\rangle_{Aab}+|666\rangle_{Aab}),\\
&|\phi'_{i}\rangle=|\phi_{i}\rangle|66\rangle_{ab}, i=11,12, \ldots, 17.\\
\end{split}
\end{equation}

Then Bob makes a 11-outcome projective measurement. For outcomes $\emph{B}_1=|2\rangle_b\langle2|\otimes|1-5\rangle_B\langle1-5|$, $\emph{B}_2=|3\rangle_b\langle3|\otimes|1-2\rangle_B\langle1-2|$, $\emph{B}_3=|4\rangle_b\langle4|\otimes|1-3\rangle_B\langle1-3|$ and $\emph{B}_4=|5\rangle_b\langle5|\otimes|1-4\rangle_B\langle1-4|$, the states $|\phi'_6\rangle, |\phi'_7\rangle, |\phi'_8\rangle$ and $|\phi'_9\rangle$ can be
identified directly.

For the outcome $\emph{B}_{i+3}=|1\rangle_b\langle1|\otimes|i\rangle_B\langle i|+|i\rangle_b\langle i|\otimes|i\rangle_B\langle i|, i=2,3,4,5$, it leaves
$|\phi'_{i}\rangle$ invariant and transforms $|\phi'_{1_{i}}\rangle$ to $|1+2+3+4\rangle_C|i\rangle_B(|111\rangle_{Aab}+|iii\rangle_{Aab})~(i=2,3,4,5).$ Alice can easily identify these states, since any two orthogonal states can be distinguished.

For the outcome $\emph{B}_9=|6\rangle_b\langle6|\otimes|1-2\rangle_B\langle1-2|$, it leaves $|\phi'_{i}\rangle~(i=11,12,13,14)$ invariant. Then, Charles can identify these
states by four projectors $\emph{C}_{9i}=|i\rangle_C\langle i|, i=1,2,3,4$.

For the outcome $\emph{B}_{10}=|3\rangle_b\langle3|\otimes|6\rangle_B\langle 6|+|6\rangle_b\langle 6|\otimes|5\rangle_B\langle 5|$, it leaves $|\phi'_{10}\rangle$,
$|\phi'_{16}\rangle$ invariant and transforms $|\phi'_{1}\rangle$ to $|1+2+3+4\rangle_C|5\rangle_B(|333\rangle_{Aab}+|666\rangle_{Aab}).$ Then, Charles uses the projector
$\emph{C}_{10,1}=|2-3\rangle_C\langle 2-3|$ to identify $|\phi'_{16}\rangle$. When Charles uses the projector
$\emph{C}_{10,2}=|4\rangle_C\langle 4|,$ it leaves $|\phi'_{10}\rangle$ invariant and transforms $|\phi'_{1}\rangle$ to $|4\rangle_C|5\rangle_B(|333\rangle_{Aab}+|666\rangle_{Aab}).$ Alice also can distinguish them, since they are two orthogonal states in Alice's Hilbert space.

For the last outcome $\emph{B}_{11}=|6\rangle_b\langle 6|\otimes|4\rangle_B\langle 4|$, it leaves $|\phi'_{15}\rangle$, $|\phi'_{17}\rangle$ invariant and transforms $|\phi'_{1}\rangle$ to $|1+2+3+4\rangle_C|4\rangle_B|666\rangle_{Aab}.$ Then, Charles can distinguish these states by the projectors $\emph{C}_{11,1}=|1-2\rangle_C\langle 1-2|,$ $\emph{C}_{11,2}=|1+2\rangle_C\langle 1+2|,$
$\emph{C}_{11,3}=|3-4\rangle_C\langle 3-4|,$ and $\emph{C}_{11,4}=|3+4\rangle_C\langle 3+4|.$

Therefore, we have succeeded in distinguishing the states (6) by LOCC using entanglement as a resource.\qed

\bigskip
In the space $\mathbb{C}^{n_1}\otimes\mathbb{C}^{n_2}\otimes\mathbb{C}^{n_3}~(4\leq n_1\leq n_2\leq n_{3}),$ there exist $2(n_1+n_3)-3$ orthogonal product states which are LOCC indistinguishable \cite{Wang17} and collectively denoted by $\mathcal{G}$.

The set $\mathcal{G}$ can be divided into three parts: $\mathcal{G}$=\{$|\phi\rangle\}\cup \mathcal{T}\cup\mathcal{R}$. The state $|\phi\rangle=|1+2+\cdots+n_1\rangle_C|1+2+\cdots+n_2\rangle_B|1+2+\cdots+n_3\rangle_A$. The subset $\mathcal{T}$ consists of $|\phi_i\rangle=|n_1\rangle\otimes|\psi_i\rangle$, $i=2,3,\ldots ,2n_3-1$, where $\{|\psi_i\rangle\}$ is the set of the states constructed in $\mathbb{C}^{n_2}\otimes\mathbb{C}^{n_3}$ using \cite[Theorem 1]{Wang17} (except the stopper state).

To describe the subset $\mathcal{R}$, we consider three cases: (a) $n_2<n_3$ and $n_3-n_2$ is odd; (b) $n_2<n_3$ and $n_3-n_2$ is even; (c) $n_2=n_3$.
Let $\{|i'\rangle\}_{i'=1}^{n_2}$ be another base in the second factor (Bob's).
By relabeling, one can such that $i'=i$ for all $i'$ in case (a);
 $(n_2-1)'=n_2,$ $n_2'=n_2-1$, $i'=i$ for the other $i'$ in case (b); and for case (c), $2'=n_2-1,$ $(n_2-1)'=2$, $i'=i$ for the other $i'$.
Let $\mathcal{R}=H\cup V$, here $H=\{|i\rangle|1'-2'\rangle|n_3\rangle \ \big| \ 1 \leq i\leq n_1\},$
and $V=\{|i-(i+1)\rangle|(n_2-\delta_{(n_1-i)})'\rangle|n_3\rangle \ \big| 1 \  \leq i \leq n_1-1\},$
where $\delta_i=\frac{1}{2}(1+(-1)^i)$.

\bigskip
\noindent{\bf Theorem \ 3.}
Over the space $\mathbb{C}^{n_1}\otimes\mathbb{C}^{n_2}\otimes\mathbb{C}^{n_3}~(4\leq n_1\leq n_2\leq n_{3}),$ it is sufficient to perfectly distinguish the set $\mathcal{G}$ of the orthogonal product states by LOCC using a maximally entangled state.

\bigskip
\noindent{\emph{ Proof:}} 
 We only consider
the case (a) to show the idea, as the other cases can be dealt with similarly.
We assume the first, second and third system belong to Charles, Bob, and Alice respectively. Similar to the previous example of the tripartite system, Alice and Bob share the $\mathbb{C}^{n_3}\otimes\mathbb{C}^{n_3}$ maximally entangled state $$|\psi\rangle_{ab}=|11\rangle_{ab}+|22\rangle_{ab}+\cdots+|{n_3}{n_3}\rangle_{ab}.$$
And then Alice performs a projector $$\emph{A}_1=|11\rangle_{aA}\langle11|+|22\rangle_{aA}\langle22|+\cdots+|{n_3}{n_3}\rangle_{aA}\langle{n_3}{n_3}|.$$
The parties can proceed from here to distinguish the states using LOCC. As the proof of Theorem 1, Alice and Bob can distinguish the states in the set
$\mathcal{T}$ except the state $|n_1\rangle_C\otimes |n_2\rangle_B\otimes |3-(n_2+1)\rangle_A$ and the state $|n_1\rangle_C\otimes |1-2\rangle_B\otimes |n_3\rangle_A(\in\mathcal{R})$. In addition, the states in the set $\mathcal{R}$ can be distinguished by Bob and Charles. Finally, the last state $|n_1\rangle_C\otimes |n_2\rangle_B\otimes |3-(n_2+1)\rangle_A$ also can be distinguished by the three parties with LOCC.

Therefore, for a general tripartite system, using a $\mathbb{C}^{n_3}\otimes\mathbb{C}^{n_3}$ maximally entangled state is sufficient to perfectly distinguish the set $\mathcal{G}$ of orthogonal product states by LOCC.\qed

\bigskip
We now generalize the method to the general odd partite quantum systems. It is known \cite{Wang17} that
there are $2(n_1+n_3+\cdots+n_{2k+1}-k)+1$ LOCC indistinguishable product states in
the space $\mathbb{C}^{n_1}\otimes\mathbb{C}^{n_2}\otimes\cdots\otimes\mathbb{C}^{n_{2k}}\otimes\mathbb{C}^{n_{2k+1}}$ $(4\leq n_1\leq n_2\leq\cdots\leq n_{2k+1})$. They are explicitly given as follows.

Let $|\psi_{i_1}\rangle_1$ $(i_1=1,2,\ldots,2(n_1+n_3)-4)$ be the product states (except the stopper state) constructed in
\cite[Theorem 3]{Wang17} for the general tripartite system. The corresponding systems are assigned to 1st Charles, Bob and Alice-system respectively.
Let $|\psi_{i_s}\rangle_s$, $i_s=1,2,\ldots,2n_{2s+1}-2$ be the product states (except the stopper state) given in
 \cite[Theorem 1]{Wang17} for the general bipartite system. The corresponding systems are designated as the s-th Alice and s-th Bob system for each integer $2\leq s\leq k$.

The set $\mathcal{S'}$ of product states in a multipartite quantum system is then the union of a stopper state $|\psi\rangle$ and
the following sets.
{\small
$$
\begin{array}{rcl}
  \mathcal{S'}_1 & = & \{|\psi_{i_1}\rangle_1|11\rangle_2\cdots|11\rangle_k\  \big| \  i_1=1,2,\ldots,2(n_1+n_3)-4\}, \\
  \mathcal{S'}_2 & = & \{|111\rangle_1|\psi_{i_2}\rangle_2|11\rangle_3\cdots|11\rangle_k  \big| \  i_2=1,2,\ldots,2(n_5)-2\}, \\
                & \vdots &                                                       \\
  \mathcal{S'}_k & = & \{|111\rangle_1\cdots|11\rangle_{k-1}|\psi_{i_k}\rangle_k  \big| \  i_k=1,2,\ldots,2(n_{2k+1})-2\},
\end{array}
$$
}
where $|\psi\rangle=|\psi_1\rangle|\psi_2\rangle\ldots|\psi_{2k+1}\rangle$ and $|\psi_i\rangle=|1\rangle+|2\rangle+\cdots+|n_i\rangle$
for $2\leq i\leq 2k+1$.

As the general odd partite quantum systems are composed of the tripartite subsystem and
even-partite subsystem, the local distinguishability
for odd-partite case is transformed into those for the tripartite system and even-partite system. Let $|\psi\rangle$ = $\otimes_{s=1}^k|\psi\rangle_{s}$ be an entangled state in $\mathbb{C}^{n_1}\otimes\mathbb{C}^{n_2}\otimes\cdots\otimes\mathbb{C}^{n_{2k}}\otimes\mathbb{C}^{n_{2k+1}}$, where $|\psi\rangle_{s}$ is a $\mathbb{C}^{n_{2s+1}}\otimes\mathbb{C}^{n_{2s+1}}$ maximally entangled state for $s = 1,2, \ldots, k.$
It follows from Theorem 2 and Theorem 3 that if the parties share the entangled state $|\psi\rangle$ we have the following result.

\bigskip
\noindent{\bf Theorem \ 4.}
 Over the space $\mathbb{C}^{n_1}\otimes\mathbb{C}^{n_2}\otimes\cdots\otimes\mathbb{C}^{n_{2k}}\otimes\mathbb{C}^{n_{2k+1}}$, it is sufficient to perfectly distinguish the set $\mathcal{S'}$ of the orthogonal product states by LOCC using entanglement as a resource.

\section {Conclusions}
Recently, much attention has been given to the problem of
distinguishing bipartite orthogonal product states by LOCC with entanglement. In this paper, we have studied the case of multipartite systems
and show that entanglement can also be used as a resource to locally distinguish the sets of multipartite product states.
We have demonstrated how to come up with the entangled state to distinguish the sets of multipartite product states in both the even and odd partite systems.
Our results provided are expected to help us understand the distinguishability of multipartite product states by LOCC.

\end{document}